\documentclass[12pt,preprint]{aastex}

\shorttitle{Search for high-energy muon neutrinos with the {IceCube} neutrino telescope}
\shortauthors{R.~Abbasi et al.} 

\usepackage{amsmath}

\newcommand{\tightlist}{
 \begin{list}{$\bullet$}
  { \setlength{\itemsep}{0pt}
     \setlength{\parsep}{3pt}
     \setlength{\topsep}{0pt}
     \setlength{\partopsep}{0pt}
     \setlength{\leftmargin}{1.5em}
     \setlength{\labelwidth}{1em}
     \setlength{\labelsep}{0.5em} } }

\newcommand{\listend}{
  \end{list}  }

\begin{document}

\title{Search for high-energy muon neutrinos from the ``naked-eye'' GRB\,080319B with the {IceCube} neutrino
	telescope}



\author{
IceCube Collaboration:
R.~Abbasi\altaffilmark{1},
Y.~Abdou\altaffilmark{2},
M.~Ackermann\altaffilmark{3},
J.~Adams\altaffilmark{4},
M.~Ahlers\altaffilmark{5},
K.~Andeen\altaffilmark{1},
J.~Auffenberg\altaffilmark{6},
X.~Bai\altaffilmark{7},
M.~Baker\altaffilmark{1},
S.~W.~Barwick\altaffilmark{8}
% ...
}
\altaffiltext{1}{Dept.~of Physics, University of Wisconsin, Madison, WI 53706, USA}
\altaffiltext{2}{Dept.~of Subatomic and Radiation Physics, University of Gent, B-9000 Gent, Belgium}
\altaffiltext{3}{DESY, D-15735 Zeuthen, Germany}
\altaffiltext{4}{Dept.~of Physics and Astronomy, University of Canterbury, Private Bag 4800, Christchurch, New Zealand}
\altaffiltext{5}{Dept.~of Physics, University of Oxford, 1 Keble Road, Oxford OX1 3NP, UK}
\altaffiltext{6}{Dept.~of Physics, University of Wuppertal, D-42119 Wuppertal, Germany}
\altaffiltext{7}{Bartol Research Institute and Department of Physics and Astronomy, University of Delaware, Newark, DE 19716, USA}
\altaffiltext{8}{Dept.~of Physics and Astronomy, University of California, Irvine, CA 92697, USA}
% ...


\begin{abstract}
We report on a search with the IceCube detector for high-energy muon
neutrinos from GRB\,080319B, one of the brightest gamma-ray bursts
(GRBs) ever observed. The fireball model predicts that a mean of 0.1
events should be detected by IceCube for a bulk Lorentz boost of the
jet of 300. In both the direct on-time window of 66\,s and an extended
window of about 300\,s around the GRB, no excess was found above
background. The 90\% CL upper limit on the number of track-like
events from the GRB is 2.7, corresponding to a muon neutrino fluence
limit of $9.5\times 10^{-3}\,\mathrm{erg}\,\mathrm{cm}^{-2}$ in the
energy range between 120\,TeV and 2.2\,PeV, which contains 90\% of the
expected events.
\end{abstract}



\section{Introduction}

Long-duration gamma-ray bursts (GRBs) are thought to originate from
the collapse of a massive star into a black hole thereby releasing a
huge amount of energy in $\gamma$-rays into the surrounding
medium. Assuming an isotropic emission of these $\gamma$-rays, the
measured fluxes yield an isotropic equivalent energy of ${\cal
O}(10^{52}$--$10^{53}$\,erg$)$. However, the actual released energy
can be significantly lower if the $\gamma$-rays are only emitted
within a small cone (jet) as suggested by the observation of
signatures for jet breaks in some X-ray spectra. Apart from being
amongst the most violent events in the universe, GRBs also belong to
the few plausible source candidates for ultra-high-energy cosmic
rays. Though our knowledge about GRBs has greatly increased in recent
years, their exact nature, the way in which particles are accelerated,
and the composition and generation of the jets formed from material
accreted onto the black hole are still not fully understood. The
observation of high-energy neutrinos from GRBs would be a smoking gun
evidence for the acceleration of hadrons in the jets and hence for the
connection between GRBs and extra-Galactic cosmic rays. 

In the fireball model \citep{apj:405:278}, neutrinos of energy ${\cal
O}(10^{14}\,\mathrm{eV})$ are produced in the decay of charged pions
generated in the interaction of accelerated protons of energy ${\cal
O}(10^{15}\,\mathrm{eV}$) with keV--MeV photons via the $\Delta^+$
resonance \citep{prl:78:2292}. Both synchrotron and inverse
Compton\footnote{Production of $\gamma$-rays through inverse Compton
emission for bursts with low optical luminosity (majority of bursts)
is actually disfavored according to \citet{mnras:393:1107} as it leads
to a very high energy $\gamma$-ray component in the TeV range which
would carry much more energy than the observed prompt $\gamma$-ray
emission resulting in an ``energy crisis'' with most current
progenitor models.}  emission from accelerated electrons have been
proposed as the mechanism for the production of these photons which
form the $\gamma$-ray signal measured by satellites. The particle
acceleration is thought to occur in internal shocks
\citep{apj:395:l83,apj:430:l93,apj:485:270} yielding $E^{-2}$ spectra
for protons and electrons as typically expected in \emph{Fermi}
acceleration
\citep{prl:78:2292}. The energy in protons (normalization of the
proton spectrum) is usually quoted in relation to the energy in
electrons which is linked to the energy in $\gamma$-ray photons
through the synchrotron and inverse Compton energy-loss mechanisms. In
the pion decay neutrinos are produced with the flavor ratios
($\nu_e$:$\nu_\mu$:$\nu_\tau$) = (1:2:0)
\footnote{Here and throughout the rest of the paper, $\nu$ denotes both
neutrinos and antineutrinos.}, changing to (1:1:1) at the Earth due to
oscillations\footnote{\citet{prl:95:181101} showed that above a
certain energy (typically ${\cal O}( 10\,\mathrm{PeV})$) the ratio
changes to (1:1.8:1.8) due to cooling energy losses of the muons
producing the neutrinos.}. Their fluence is proportional to the
fluence in $\gamma$-rays.

The first calculations of the expected neutrino flux from GRBs
\citep{prl:78:2292,apj:521:928} used average GRB parameters and the
GRB rate measured by BATSE to determine an all-sky neutrino flux from
the GRB population. This so-called Waxman--Bahcall GRB flux or similar
GRB fluxes have been probed with the AMANDA-II neutrino telescope
\citep{apj:664:397,apj:674:357} with negative results. These
fluxes will be detectable by next-generation neutrino telescopes like
IceCube with an instrumented volume of $\gtrsim
1\,\mathrm{km}^3$. However, the average flux for a single burst
derived in these models is very small even for km$^3$
detectors. Nevertheless, as the expected neutrino flux can actually
vary by orders of magnitude between GRBs due to fluctuations in the
burst parameters \citep{pr:d62:093015,app:25:118}, the detection of
extremely bright GRBs in neutrinos does seem possible, albeit
requiring at least a km$^3$-size detector, as, for example, was
demonstrated in the analysis of GRB\,030329 with the AMANDA detector
\citep{proc:icrc05:stamatikos:1}.

On 2008 March 19, at 06:12:49 UT GRB\,080319B \citep{nature:455:183}
was detected by the \emph{Swift} \citep{ssr:120:165} and Konus-Wind
\citep{wind:homepage} satellites at  $\mathrm{R.A.} =
217^\circ\!\!.\,9$ and $\mathrm{decl.} = 36^\circ\!\!.\,3$. It was the optically
brightest GRB ever observed and with a peak magnitude of 5.3 even
visible to the naked eye for a short period of time, despite the fact
that the corresponding redshift was about 0.9. It is also one of best
measured GRBs with optical wide-field observations covering the whole
duration of the explosion
\citep{gcn:circular:7445} and with many triggered follow-up
observations spanning the electromagnetic energy spectrum from radio
to $\gamma$-rays (\citet{nature:455:183} and references therein).

\section{Neutrino spectrum calculation}\label{sec:nuspec}
We calculate the expected prompt neutrino spectrum for GRB\,080319B in
the internal shock scenario of the fireball model following
\citet{app:20:429} which is based on \citet{prl:78:2292}. This model
allows for the easy incorporation of many measured parameters of
GRB\,080319B (in \citet{prl:78:2292} average GRB parameters are used)
which is crucial when investigating a GRB that deviates largely from
the average GRB. Other models like
\citet{pr:d73:063002}\footnote{This model is actually similar to
\citet{prl:78:2292} but uses Monte Carlo simulation to calculate the
photomeson production in $p\gamma$ interactions and takes the
synchrotron losses of mesons and protons into account.} or
\citet{prl:90:241103} do not provide this possibility without
obtaining the actual simulation code and are therefore not considered
here.

\begin{figure}[t]
\epsscale{.7}
\plotone{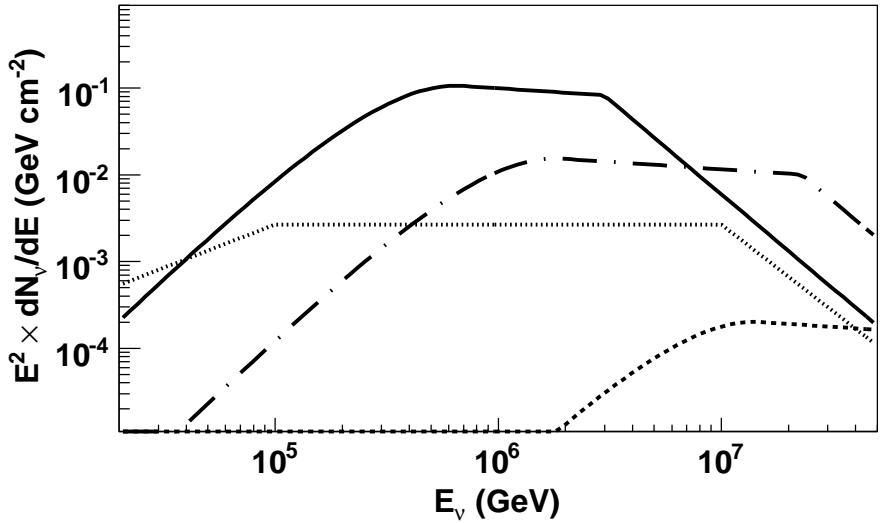}
\caption{Calculated muon neutrino spectrum for different bulk
Lorentz boost factors $\Gamma_\mathrm{jet}$ of the GRB jet:
$\Gamma_\mathrm{jet} = 300$ (solid line), $\Gamma_\mathrm{jet} = 500$
(dash-dotted line), and $\Gamma_\mathrm{jet} = 1400$ (dotted line). For
comparison, the average Waxman--Bahcall GRB fluence for a single GRB is
also shown (fine-dotted line).\label{fig:nuspec}}
\end{figure}

For reference, we list all formulae used in our calculations in
Appendix \ref{app:nu_spec}. One of the major inputs to the model is the
keV--MeV $\gamma$-ray spectrum recorded by the satellites. In contrast
to \citet{app:20:429} where a broken power law was used, we
parameterize the $\gamma$-ray spectrum with a Band function
(\citet{apj:413:281}; Equation~(\ref{eq:gamma_spec})). The function
parameters obtained from a fit to the time integrated Konus-Wind
spectrum are taken from
\citet{nature:455:183} (suppl. information) and are listed in
Table~\ref{tab1}. The table also contains additional parameters with
their values required by the model. Not all of them are measured
or even well known. For the jet parameters $\epsilon_e$ (fraction of
jet energy in electrons), $\epsilon_B$ (fraction of jet energy in
magnetic field) and $f_e$ (ratio between energy in electrons and
protons) typical values of 0.1 are used \citep{pr:458:173}. The
variability of the $\gamma$-ray flux $t_\mathrm{var}$, which is used
as a measure for the time between the emission of two consecutive
shells, was analyzed in \citet{proc:aipcp:1065:259}. They find an
initial time scale of 0.1\,s which increased to 0.7\,s in the course
of the emission. However, their analysis was limited to the energy
range between 15 and 150\,keV, where for high energies (100--150\,keV)
the dominant time scale was significantly shorter (0.05\,s). By
contrast, muons reconstructed in the IceCube detector are mainly
produced by neutrinos near the first break energy
(Figures~\ref{fig:nuspec} and \ref{fig:effArea}(b)) which originate from
proton interactions with $\gamma$-rays around 500\,keV (break energy
in $\gamma$-ray spectrum). Due to the large gap between 150 and
500\,keV and the lack of an extrapolation method we use a typical
value of $t_\mathrm{var} = 0.01$\,s \citep{pr:458:173} in our
calculations.

\begin{deluxetable}{lcl}
\tablecaption{Fireball Model Parameters Used in the Calculation of the Neutrino
Spectrum for GRB\,080319B\label{tab1}}
\tabletypesize{\scriptsize}
\tablewidth{0pt}
\tablehead{
\colhead{Parameter} & \colhead{Value} & \colhead{Reference}
}
\startdata
$E_\gamma^\mathrm{iso}$                                 & $1.3 \times 10^{54}\,\mathrm{erg}$ 
& \citet{nature:455:183}\\ 
Burst duration                                   & 66\,s
& \citet{gcn:report:134.1}\\ 
$\Gamma_\mathrm{jet}$                            & 300, 500, 1400
& \citet{nature:455:183}; see also main text\\
$\gamma$ spec.---fluence, ${\cal F}_\gamma$ (20\,keV--7\,MeV)      & $6.23 \times 10^{-4}$\,erg\,cm$^{-2}$
& \citet{nature:455:183}\\
$\gamma$ spec.---break energy, $\epsilon_\gamma$ & 651\,keV
& \citet{nature:455:183} (suppl. information)\\
$\gamma$ spec.---1st index, $\alpha_\gamma$      & 0.83
& \citet{nature:455:183} (suppl. information)\\
$\gamma$ spec.---2nd index, $\beta_\gamma$       & 3.5
& \citet{nature:455:183} (suppl. information)\\
z                                                & 0.94
& \citet{gcn:circular:7451}\\ 
$x_\mathrm{\pi}$\tablenotemark{a}                & 0.2 
& \citet{pr:458:173}\\
$\epsilon_e$                                            & 0.1 
& \citet{pr:458:173}\\
$\epsilon_B$                                            & 0.1 
& \citet{pr:458:173}\\
$f_e$                                            & 0.1 
& \citet{pr:458:173}\\
$t_\mathrm{var}$                                 & 0.01\,s 
& \citet{pr:458:173}\\
\enddata
\tablenotetext{a}{Fraction of proton energy going into pion in a
single $p\gamma$ interaction.} 
\end{deluxetable}

The neutrino spectrum is parameterized as a Band function with a
broken power law at high energies (Equation~(\ref{eq:nu_spec})), where
the latter describes the steepening of the spectrum due to synchrotron
losses of pions and muons. The energy in neutrinos, i.e., the
normalization of the spectrum, is proportional to $f_e^{-1}$ and the
energy in $\gamma$-ray photons (Equation~(\ref{eq:nuIntegral})). The
$\Gamma_\mathrm{jet}$ factor, which enters the model equations to the
second and fourth power (Equations~(\ref{eq:epsilon1}),
(\ref{eq:epsilon2}), and (\ref{eq:nint})), has a large impact on the
normalization of the neutrino fluence. With increasing
$\Gamma_\mathrm{jet}$, the shell collisions occur at larger distances
from the black hole where the photon field and hence the target
density for the pion production is smaller. The value of
$\Gamma_\mathrm{jet}$ can be estimated from the fact that the source
has to be transparent for $\gamma$-rays near the maximum $\gamma$-ray
energy produced \citep{app:20:429}. Using the parameters given in
Table~\ref{tab1} and a maximum $\gamma$-ray energy of 100\,MeV yields
$\Gamma_\mathrm{jet}
\approx 300$. In \citet{nature:455:183}, the authors argue that the
exceptional brightness of the optical flash in GRB 080319B implies
that the self-absorption frequency cannot be far above the optical
band and they determine $\Gamma_\mathrm{jet}$ to lie between 300 and
1400. \citet{arXiv:0904.1797} obtain $\Gamma_\mathrm{jet} \approx 300$
from an extrapolation of the late-time evolution of the
afterglow. Using a synchrotron self-Compton model,
\citet{mnrasl:391:l19} find a $\Gamma_\mathrm{jet}$ factor of
$\sim\!500$. For the calculation of the neutrino spectrum, we adopt the
optimistic case with $\Gamma_\mathrm{jet} = 300$ which is displayed in
Figure~\ref{fig:nuspec} as the solid line. An increase of the
$\Gamma_\mathrm{jet}$ factor to 500 (1400) decreases the neutrino
fluence by about a factor 10 ($10^{3}$) and shifts it to higher
energies (Figure~\ref{fig:nuspec} dashed and dotted lines). Muon cooling
\citep{prl:95:181101} affects the neutrino spectrum only at energies
above $\sim\!20$\,PeV ($\Gamma_\mathrm{jet} = 300$) and is therefore
negligible for our analysis.

\section{Analysis of IceCube data}
IceCube \citep{app:26:155}, the successor of the AMANDA experiment and
the first next-generation neutrino telescope, is currently being
installed in the deep ice at the geographic South Pole. Its final
configuration will instrument a volume of about $1\,\mathrm{km}^3$ of
clear ice in depths between 1450\,m and 2450\,m. Neutrinos are
reconstructed by detecting the Cherenkov light from charged secondary
particles, which are produced in interactions of the neutrinos with
the nuclei in the ice or bedrock. The optical sensors consist of
photomultipliers housed in pressure-resistant glass spheres (digital
optical modules (DOMs); \citet{nim:a601:294}) which are mounted on
vertical strings. Each string carries 60 DOMs with the final
detector containing 80 such strings. Physics data taking with IceCube
started in 2006 with nine strings installed. At the beginning of 2007, the
detector was enlarged to 22 strings. Since 2009 April, IceCube has been
running with 59 strings. The completion of the detector construction
is planned for the year 2011.

\begin{figure}[t]
\epsscale{.6}
\plotone{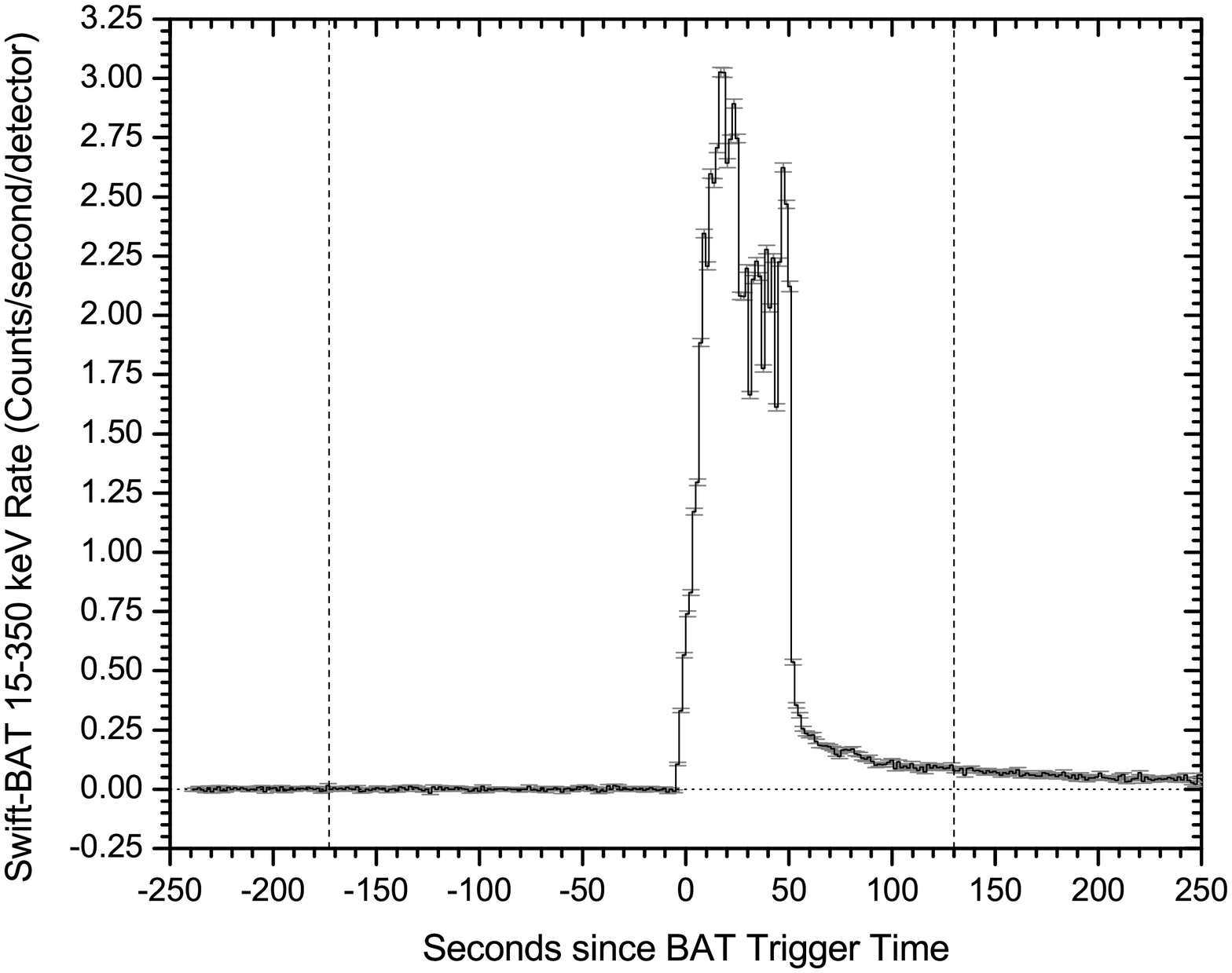}
\caption{$\gamma$-ray emission from GRB\,080319B as measured by
\emph{Swift}-BAT \citep{privcom:stamatikos:2008:1} with $T_0 =$ 06:12:49 UT.
The dashed vertical lines mark the time range covered by IceCube
data.\label{fig:gammacurve}}
\end{figure}

\begin{figure}[t]
\epsscale{1}
\plottwo{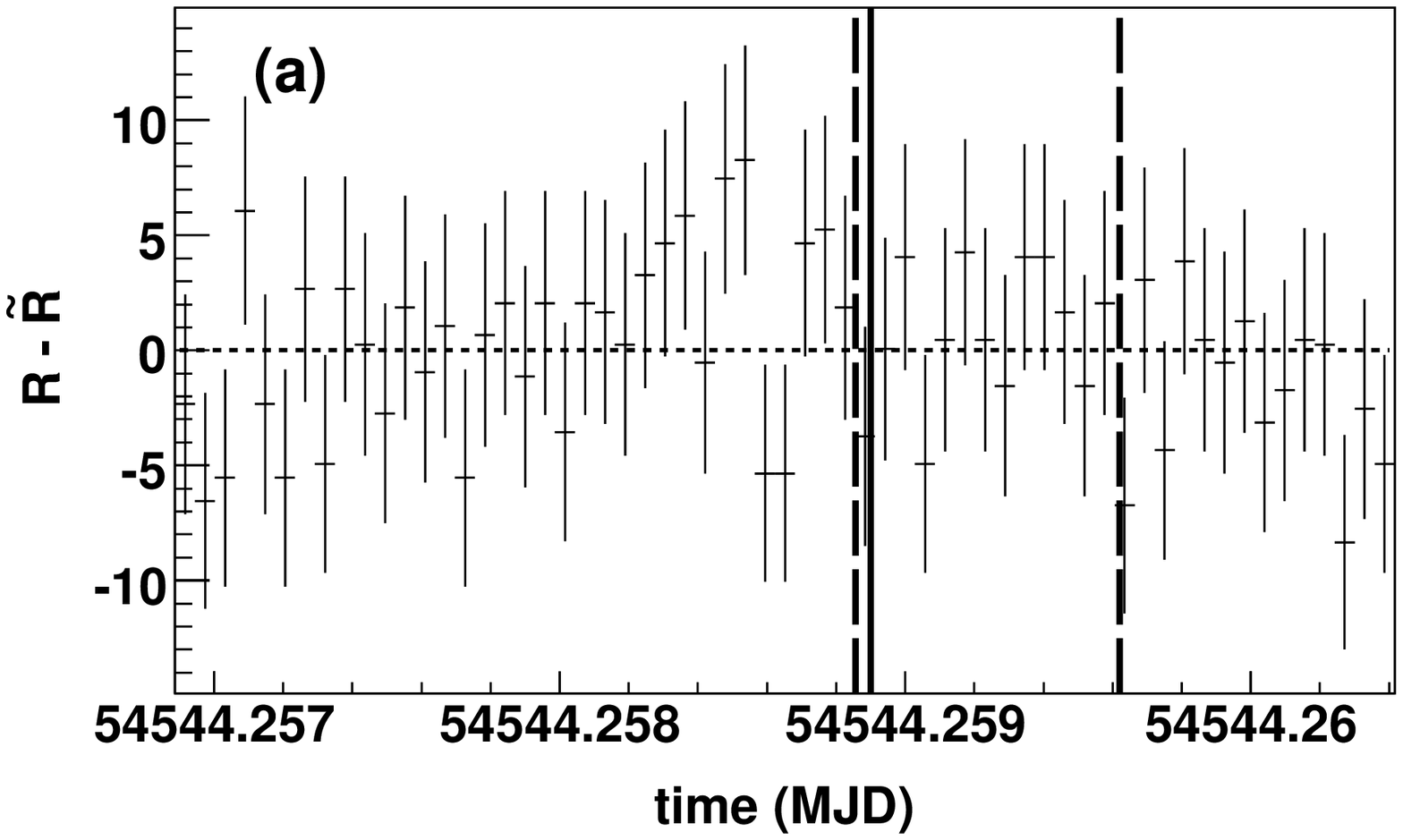}{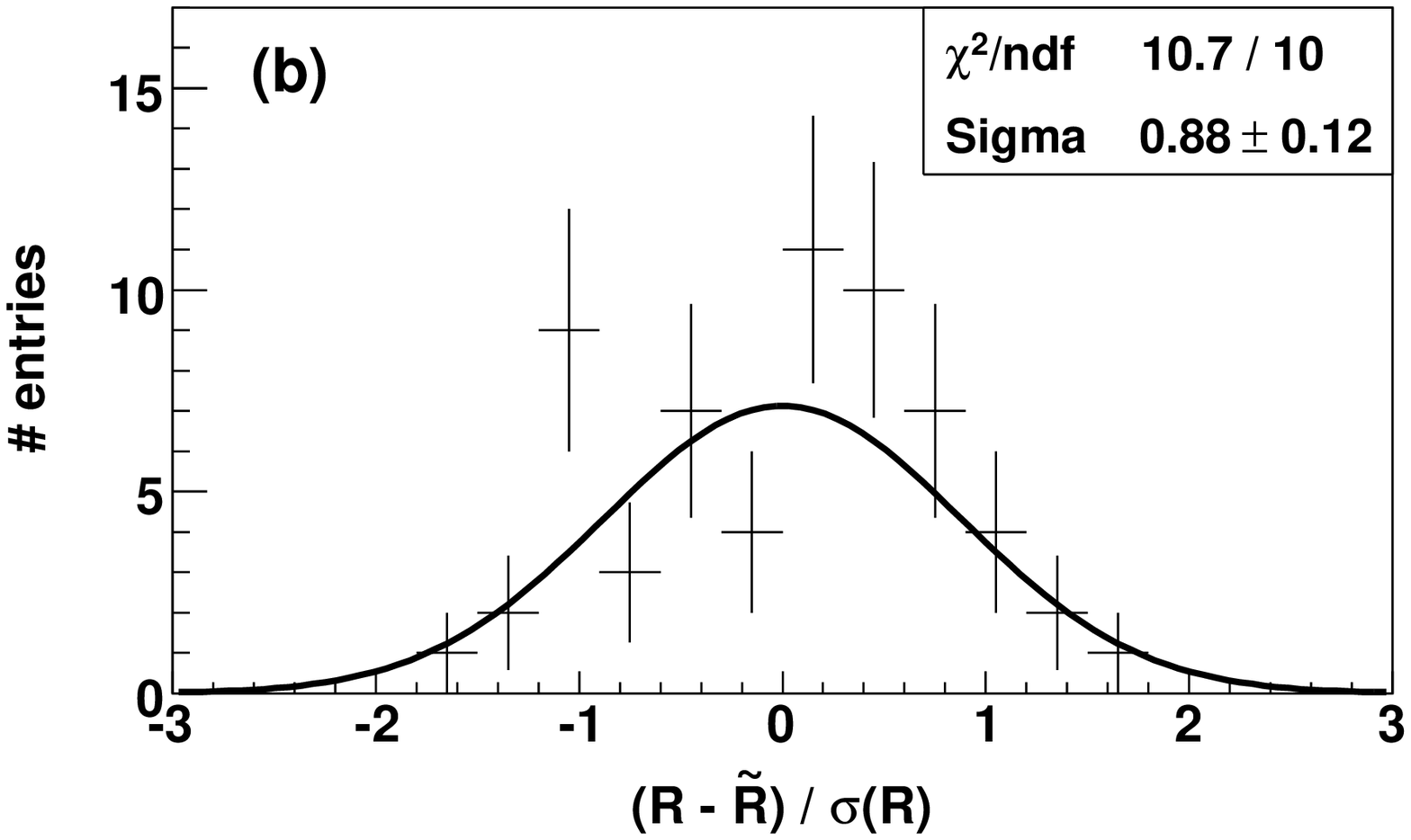}
\caption{(a) Difference between the event rate in a 5\,s bin, $R$, and
the average rate $\tilde{R}$ (calculated from all events in the 300\,s
time window shown) as a function of time at trigger level. The solid
and dashed vertical lines mark the satellite trigger time of the GRB
and the start and stop times of the measured $\gamma$-ray emission,
respectively. (b) Histogram of ($R-\tilde{R}$) divided by the
statistical errors. The line is a fit of a Gaussian distribution to
the histogram.\label{fig:stability}}
\end{figure}

\begin{deluxetable}{lp{3mm}ccp{3mm}ccl}
\tablecaption{Number of Expected Signal and Background Events at Different Cut Levels\label{tab2}}
\tabletypesize{\scriptsize}
\tablewidth{0pt}
\tablehead{
                    && \multicolumn{2}{c}{Signal}                                                   && \multicolumn{2}{c}{Background (Off-Time Data)} \\
\colhead{Cut Level} && \colhead{No. Events} & \colhead{Efficiency\tablenotemark{a}} (\%)&& \colhead{No. Events\tablenotemark{b}} & \colhead{Efficiency\tablenotemark{a}} (\%)& \colhead{Comment}
}
\startdata
Trigger      &&$0.24$	&  100 && $1.2\times10^{-1}$  & 100   &See Section~\ref{sec:data}\\
Quality      &&$0.14$	&   58 && $1.4\times10^{-3}$  & 1.2   &See Equation~(\ref{eq:qualityCuts})\\
Final        &&$0.10$	&   41 && $1.7\times10^{-5}$  & 0.014 &See Equation~(\ref{eq:finalCuts})\\
\enddata
\tablenotetext{a}{Relative to trigger level.} 
\tablenotetext{b}{In a cone with radius $5^\circ$ centered on GRB\,080319B position within 66\,s.} 
\end{deluxetable}

\subsection{Data sets, event reconstruction and selection}\label{sec:data}
The data acquisition (DAQ) system of IceCube \citep{nim:a601:294} is
based on local coincidences between neighboring DOMs in a string
within $1\,\mu$s for which the photon signal passes a threshold of 0.4
photo electrons. All data from DOMs belonging to a local coincidence
are read out and the digitized waveforms are sent to a computer farm
at the surface. In order to pass the trigger, a minimum number of eight
DOMs in local coincidences within a time window of 5000\,ns is
required. If this condition is fulfilled, the waveforms are combined to
an event and the number and arrival times of the Cherenkov photons are
extracted. For each event, an initial track is reconstructed using the
line-fit algorithm \citep{nim:a524:169}. This is a simple but fast
track reconstruction based on the measured hit times in the DOMs. We
do not consider events where only a shower is produced (e.g., in
interactions of electron neutrinos or neutral current neutrino
interactions) as only the track-like light pattern of muons allows for
a good angular resolution. Hence, this search focuses on muon
neutrinos from GRB\,080319B. Cuts on the reconstructed zenith angle
($> 70^\circ$) and number of hit DOMs ($>10$) reject downgoing
atmospheric muons and reduce the event rate down to the 117\,Hz. This
allows to apply more advanced reconstruction algorithms to the
remaining neutrino candidates. A more precise estimate of the
direction of a neutrino candidate is obtained by fitting a muon-track
hypothesis to the pattern of the recorded Cherenkov light in the
detector using a log-likelihood reconstruction method
\citep{nim:a524:169}. A fit of a paraboloid to the region around the
minimum of the log-likelihood function yields an estimate of the
uncertainty on the reconstructed direction. The absolute time of an
event is determined with a GPS clock with a precision of better than a
millisecond, which is more than sufficient for this analysis.

At the time of GRB\,080319B, IceCube was running in maintenance mode
with 9 out of 22 strings taking data. Apart from the reduced number of
strings the DAQ system had a slightly different configuration than
during normal operation. IceCube data are available in a window of
about 300\,s around the GRB (on-time data) as displayed in
Figure~\ref{fig:gammacurve}. The detector was checked for stability during this
period by plotting the rate of events passing the trigger in bins of
time, $R$, relative to the average rate $\tilde{R}$ of 117\,Hz
(Figure~\ref{fig:stability}). The variations in the event rate are compatible
with statistical fluctuations and there were no indications for
abnormal behavior of the detector during the period under
consideration.

\begin{figure}[t]
\epsscale{1}
\plotone{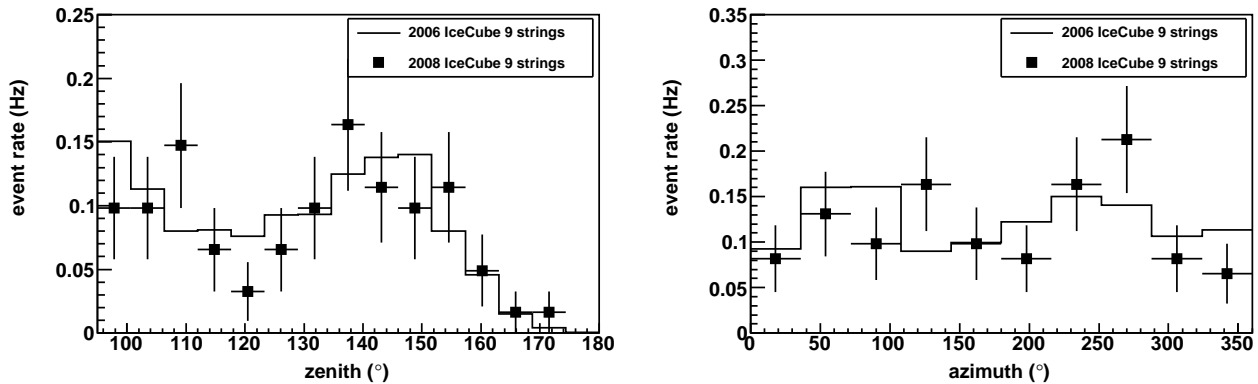}
\caption{Comparison after quality cuts (Equation~(\ref{eq:qualityCuts}))
of the background data set (2006 IceCube 9 strings) with a 1\,hr
data set taken 1 week after GRB\,080319B with equivalent DAQ
settings (2008 IceCube 9 strings).\label{fig:dataData}}
\end{figure}

In order to avoid systematic uncertainties due to inaccuracies in the
simulation when calculating the significance of a deviation from the
background-only hypothesis, the expected background in the on-time
window is determined from the observed off-time data. However, the
amount of data taken with the special DAQ configuration during the
regular maintenance runs is limited and not sufficient for a good
background estimation. Instead, we utilize the IceCube data set of the
2006 data taking period (131 days lifetime), when the detector
consisted only of the same nine strings which were taking data during
GRB\,080319B. After applying the following quality cuts, the 2006 data
set shows good agreement with the GRB\,080319B maintenance data set
(Figure~\ref{fig:dataData})
\begin{equation}
\theta_\mathrm{rec} > 90^\circ \ \ ; \ \ 
\sigma_\mathrm{dir} < 10^\circ \ \ ; \ \
\theta_\mathrm{min} > 70^\circ \ \ ; \ \
L_\mathrm{red} \le 
  \begin{cases} 
    9.0 \quad \text{for } 4 \le N_\mathrm{dir} \le 10\\
    8.0 \quad \text{for } N_\mathrm{dir} < 4
  \end{cases}
\label{eq:qualityCuts}
\end{equation}
with
\tightlist
\item $\theta_\mathrm{rec}$: reconstructed zenith angle;
\item $\sigma_\mathrm{dir}$: uncertainty on the reconstructed track direction
(quadratic average of the minor and major axis of the $1\sigma$ error
ellipse);
\item $L_\mathrm{red}$: $-\log_{10}$ of the likelihood value of the reconstructed track
divided by the number of degrees of freedom (number of hit DOMs minus
number of fit parameters). In conjunction with the selection of
upgoing tracks it has proven to be an efficient variable for
separating upgoing atmospheric neutrinos from misreconstructed
downgoing atmospheric muons. It exploits the fact that for a light pattern
originating from a downgoing muon the incorrect upgoing track
hypothesis yields rather low likelihood values;

\item $N_\mathrm{dir}$: number of photons detected within a $-15$
to $+75$\,ns time
window with respect to the expected arrival time for unscattered
photons from the muon-track hypothesis; and
\item $\theta_\mathrm{min}$: minimum of the two zenith angles from a fit of a two-track hypothesis to
the light pattern.
\listend
The $\theta_\mathrm{min}$ cut rejects events where two downgoing
muons from independent atmospheric showers pass through the detector
in quick succession and produce a light pattern that fakes an upgoing
track. The difference in the overall rates of the two samples is about
10\% which is within the statistical error of the total number of
events in the maintenance data set.

For the final analysis, we use the method described in
Section~\ref{sec:anamethod} to maximize the potential for a discovery:
the cuts on $\sigma_\mathrm{dir}$ and $N_\mathrm{dir}$ are tightened
until we reach a maximum in the probability to detect the fluence
calculated in Section~\ref{sec:nuspec} ($\Gamma_\mathrm{jet} = 300$)
with a significance of at least $4\sigma$. In order to avoid biasing
the results, only the off-time data are used and the on-time data are
kept ``blind'' during this procedure. The optimized cuts are
\begin{equation}
\theta_\mathrm{rec} > 90^\circ \ \ ; \ \ 
\sigma_\mathrm{dir} < 5^\circ \ \ ; \ \
\theta_\mathrm{min} > 70^\circ \ \ ; \ \
L_\mathrm{red} \le 
  \begin{cases} 
    9.0 \quad \text{for } 8 \le N_\mathrm{dir} \le 10\\
    8.0 \quad \text{for } N_\mathrm{dir} < 8
  \end{cases}.
\label{eq:finalCuts}
\end{equation}
After these cuts the data sample with an event rate of about
$5\times10^{-2}$\,Hz is still dominated by misreconstructed
downgoing muons (a Monte Carlo simulation of atmospheric neutrinos
yields an event rate of $2\times10^{-3}$\,Hz). For a search cone with
$5^\circ$ radius centered on the GRB position and a time window of
66\,s, a mean number of background events of $1.7\times10^{-5}$ is
expected. A summary of the event rates at different cut levels is
given in Table~\ref{tab2}.

\subsection{Monte Carlo}
Signal muon neutrinos from GRB\,080319B are generated with a Monte
Carlo and weighted according to the fluence calculated in
Section~\ref{sec:nuspec}. The Monte Carlo contains a detailed
simulation of the propagation of the muon neutrino through the Earth
and the ice using the ANIS generator \citep{cpc:172:203}. After the
interaction, the muon is traced through rock and ice taking into
account continuous and stochastic energy losses
\citep{hep-ph-0407075}. The photon signal in the DOMs is determined
from a detailed simulation
\citep{nim:a581:619} of the propagation of Cherenkov light through the
ice which includes the modeling of the changes in absorption and
scattering length with depth
\citep{jgr:111:D13203}. This is followed by a simulation
of the DOM electronics and the trigger. The DOM signals are then
processed in the same way as the real data. Background events do not
have to be simulated as these are taken from off-time data.

\begin{figure}[t]
\epsscale{1}
\plottwo{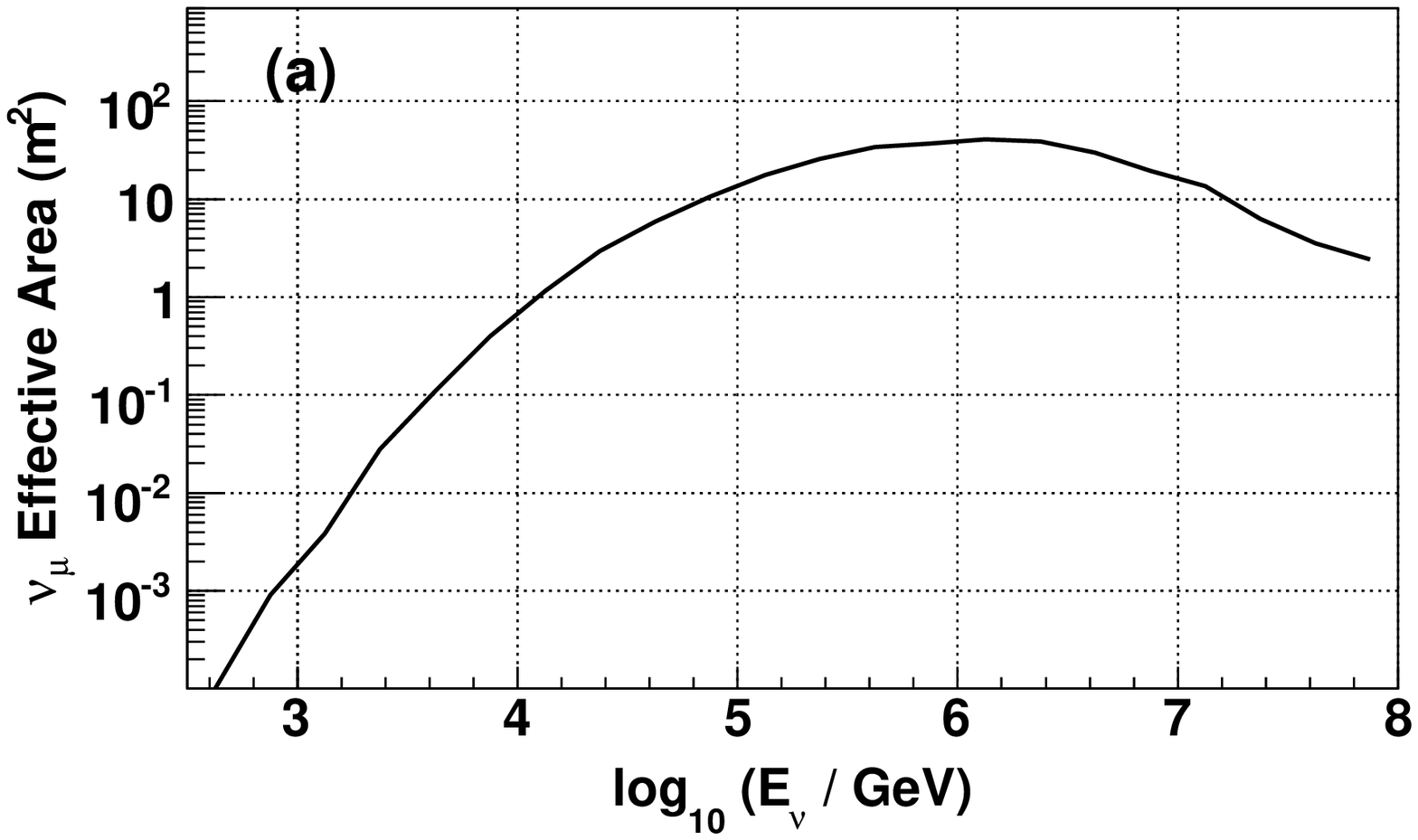}{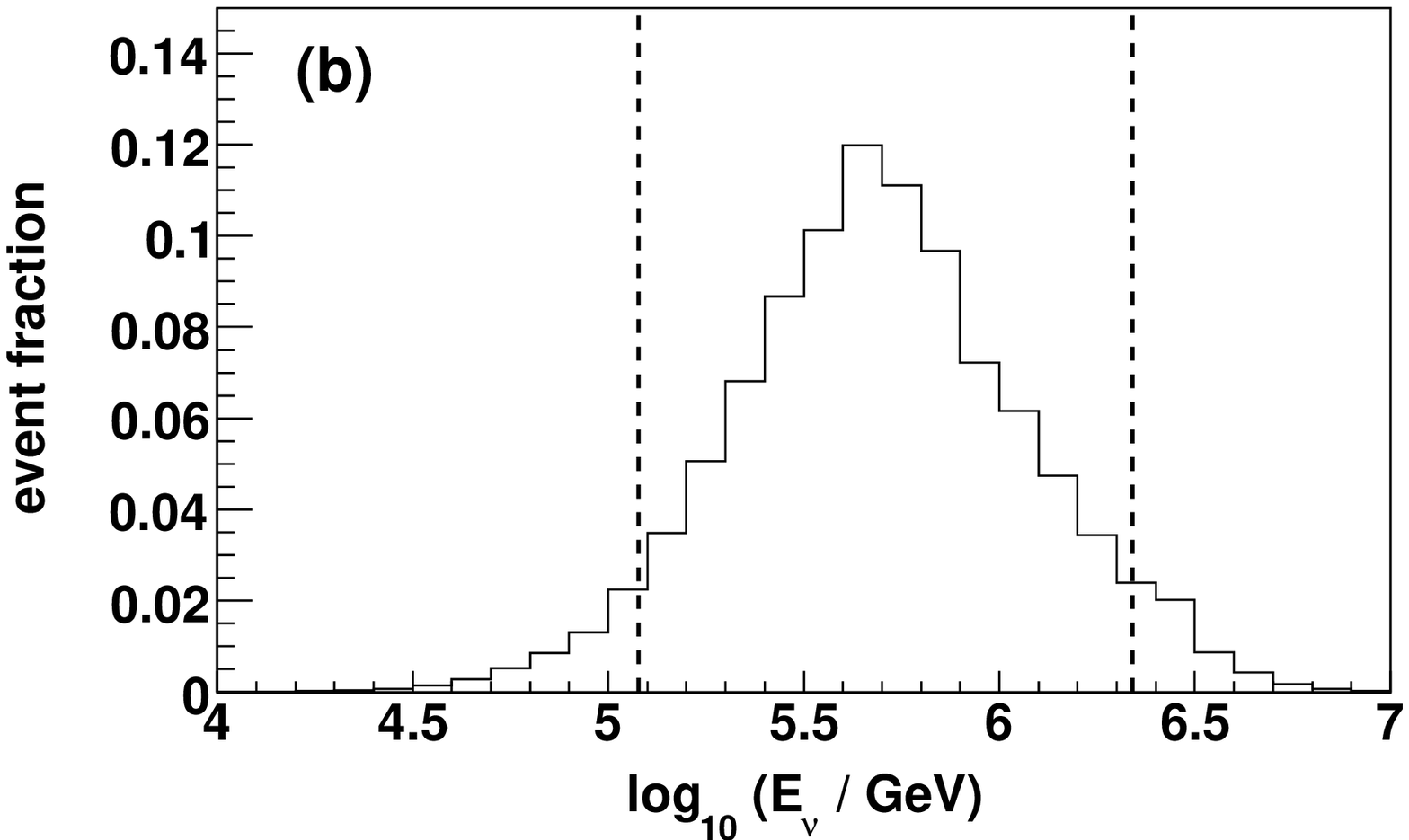}
\caption{(a) Effective area at final cut level for muon neutrinos
from the direction of GRB\,080319B as a function of the neutrino
energy. (b) Expected mean number of Monte Carlo signal events after
final cuts as a function of neutrino energy. The two vertical lines
mark the central interval containing 90\% of the
events.\label{fig:effArea}}
\end{figure}

After final cuts (Equation~(\ref{eq:finalCuts})) and for neutrinos from
the direction of the GRB (weighted according to the calculated GRB
spectrum), 90\% of all reconstructed Monte Carlo tracks are contained
within $20^\circ$ of the true direction. The median angular resolution
is $5.6^\circ$. This resolution is worse than the one usually quoted
for IceCube in its nine-string configuration for neutrino-induced
muons. The reason for this is that the geometry of the detector was
such that the reconstruction lever arm was at its shortest for the
direction of GRB\,080319B. The corresponding muon neutrino effective
area as a function of energy is displayed in
Figure~\ref{fig:effArea}(a). The expected mean number of events from the
GRB ($\Gamma_\mathrm{jet} = 300$) after final cuts is 0.1, with 90\%
of the events contained in the energy range from 120\,TeV to 2.2\,PeV
(Figure~\ref{fig:effArea}(b)). A summary of the event rates at different
cut levels is given in Table~\ref{tab2}.

\begin{figure}
\epsscale{0.8}
\plotone{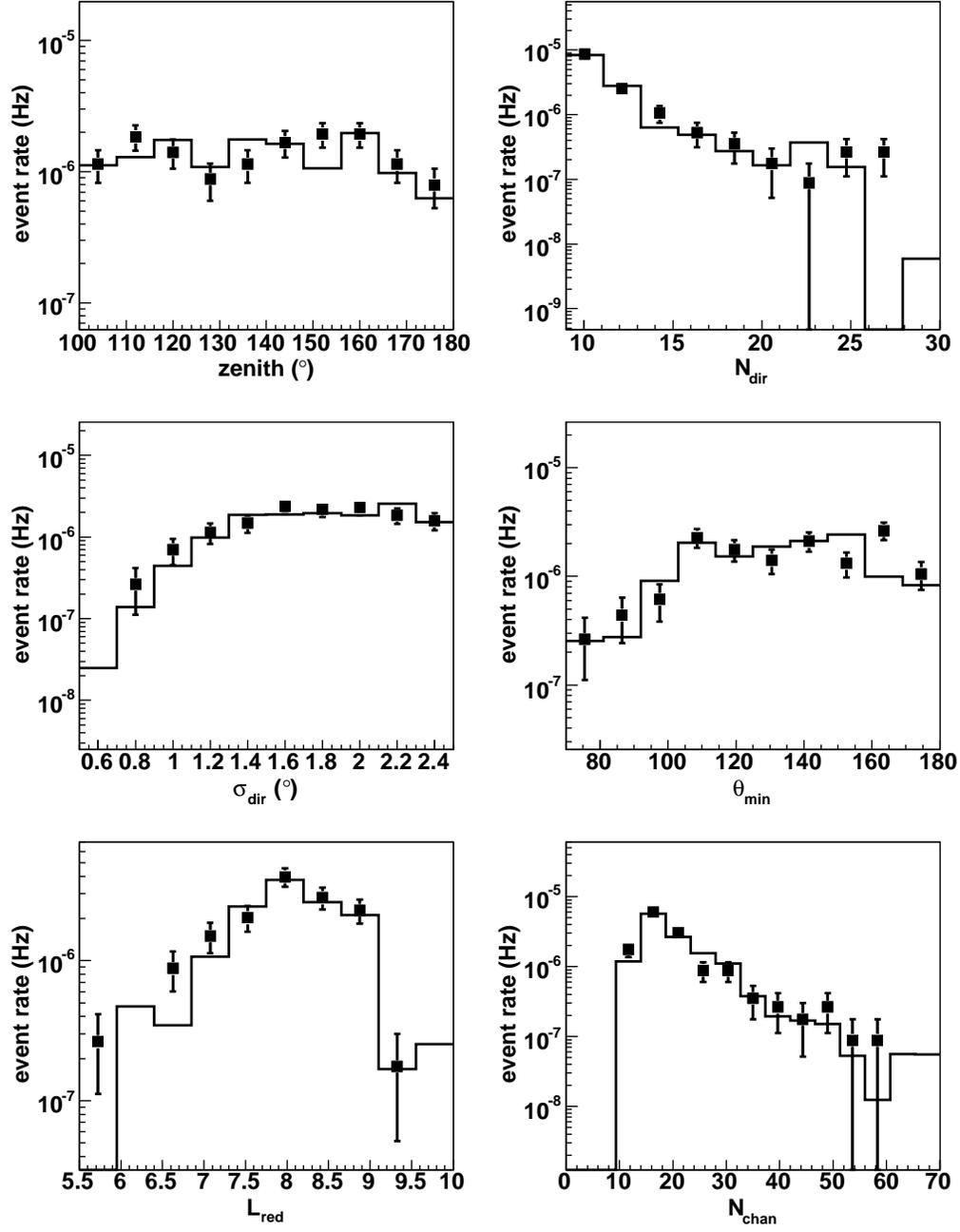}
\caption{Comparison between an atmospheric muon neutrino Monte Carlo
(solid line) and a high-purity data set of (atmospheric) muon
neutrinos (filled squares) in different
variables.\label{fig:sigDataComp}}
\end{figure}

In order to compare the signal (neutrino) Monte Carlo to data, a
high-purity (atmospheric) muon neutrino sample from the off-time data
is selected by requiring $\mathrm{zenith} > 100^\circ$,
$\sigma_\mathrm{dir} < 2^\circ\!\!.\,5$, and $N_\mathrm{dir} > 8$ (these
tight cuts are only used for this comparison and not in the following
analysis of the data). Figure~\ref{fig:sigDataComp} shows good
agreement between data and the Monte Carlo weighted to the atmospheric
muon neutrino flux of
\citet{pr:d70:023006} (Bartol spectrum).

\subsection{Unbinned likelihood analysis}\label{sec:anamethod}
The final analysis is based on data sets at final cut level
(Equation~(\ref{eq:finalCuts})). The data are analyzed using an unbinned
log-likelihood method similar to the one described in
\citet{app:29:299}. In contrast to binned methods where the event is
rejected if it lies outside the cut region (binary selection),
unbinned likelihood methods do not throw away events but use
probability density functions (PDFs) to evaluate the probability of an
event to belong to signal or background. Therefore, they are more
powerful than binned methods.

In the case of searches for neutrinos from GRBs detected by satellites,
the direction and time of reconstructed tracks are the most crucial
information to separate signal from background. Hence, both the signal,
$S(\vec{x}_i)$, and background, $B(\vec{x}_i)$, PDFs are each the
product of a time PDF and a directional PDF, where $\vec{x}_i$ denotes
both the directional and time variables. The directional signal PDF is
a two-dimensional Gaussian distribution with the two widths being the
major and minor axis of the $1\sigma$ error ellipse of the paraboloid
fit described in the previous section. The time PDF is flat over the
$\gamma$-ray emission time and falls off on both sides with a Gaussian
distribution with $\sigma = 25$\,s. The Gaussian accounts for possible
small shifts in the neutrino emission time with respect to that of the
$\gamma$-rays and prevents discontinuities in the likelihood
function. The sensitivity of the analysis depends only weakly on the
exact choice of $\sigma$, e.g., the quoted upper limit changes by less
than 2\% when doubling $\sigma$. For the directional background PDF,
the detector asymmetries in zenith and azimuth have to be taken into
account. This is accomplished by evaluating the data in the detector
coordinate system. The directional background PDF is hence derived
from the distribution of all background events after final cuts in the
zenith--azimuth plane of the detector. The time distribution of
background tracks during the GRB can be assumed to be constant
resulting in a flat time PDF.

Both PDFs are combined in an extended log-likelihood function
\citep{barlow:statistics:1989}
\begin{equation}
\ln\left({\cal L}(\langle n_s \rangle)\right) = -\langle n_s \rangle - \langle n_b \rangle + \sum_{i=1}^{N} \ln\left(
\langle n_s \rangle\,S(\vec{x}_i) + \langle n_b \rangle\,B(\vec{x}_i) \right) \ ,
\end{equation}
where the sum runs over all reconstructed tracks left after cuts with
$\vec{x}_i$ representing the PDF parameters (absolute time of the
track along with the track direction in local detector coordinates and
its estimated uncertainty). The variable $\langle n_b \rangle$ is the expected mean
number of background events, which is determined from the background
data set. The mean number of signal events, $\langle n_s \rangle$, is a free parameter
which is varied to maximize the expression
\begin{equation}
\ln\left({\cal R}(\langle n_s \rangle)\right) = \ln\left(\frac{{\cal L}(\langle n_s \rangle)}{{\cal L}(0)}\right) = -\langle n_s \rangle +
	\sum_{i=1}^N \ln \left(
	\frac{\langle n_s \rangle\,S(\vec{x}_i)}{\langle n_b \rangle\,B(\vec{x}_i)} + 1 \right)
\end{equation}
in order to obtain the best estimate for the number of signal events,
$\langle \hat{n}_s \rangle$.

\begin{figure}[t]
\epsscale{1.1}
\plottwo{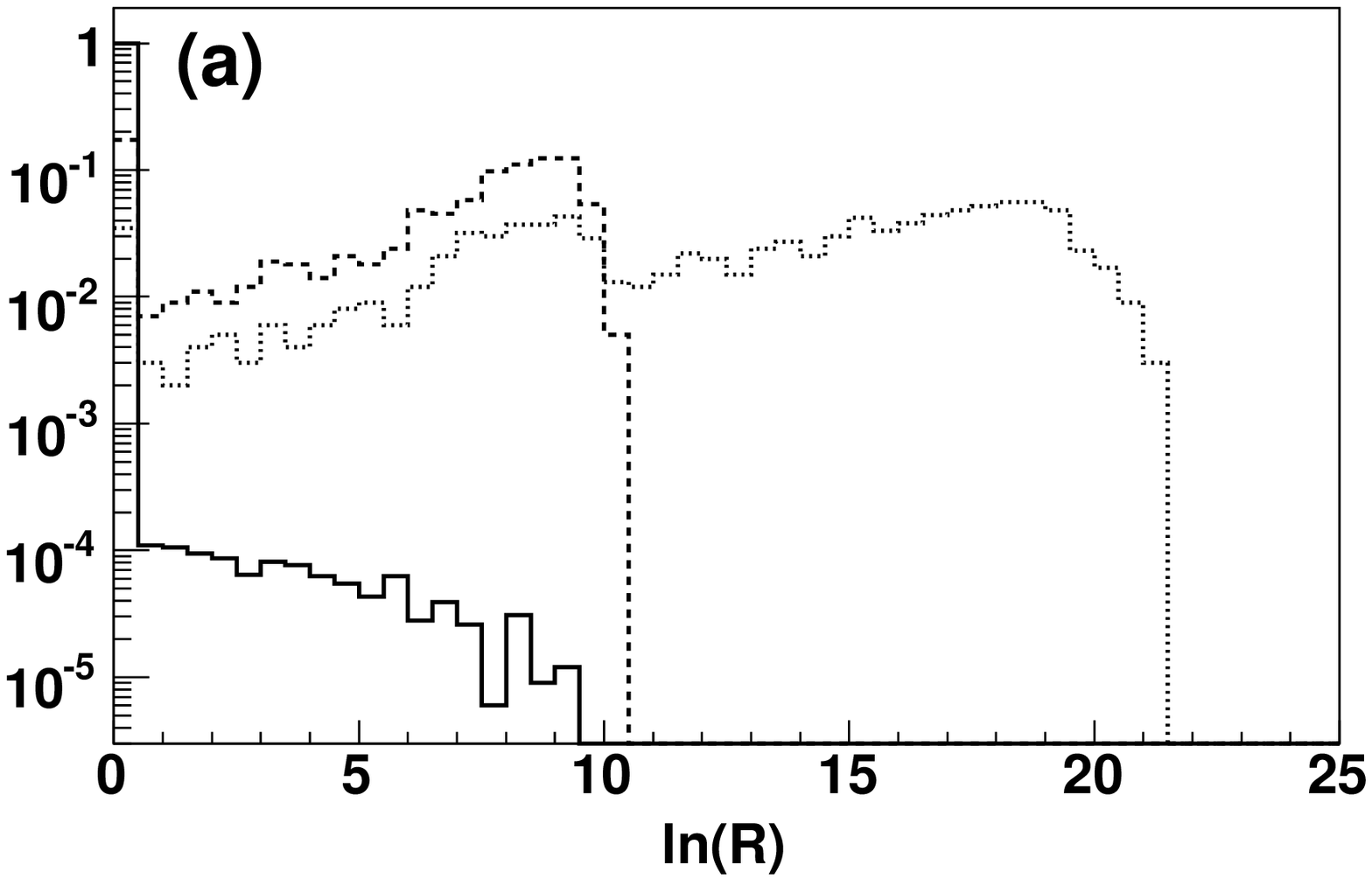}{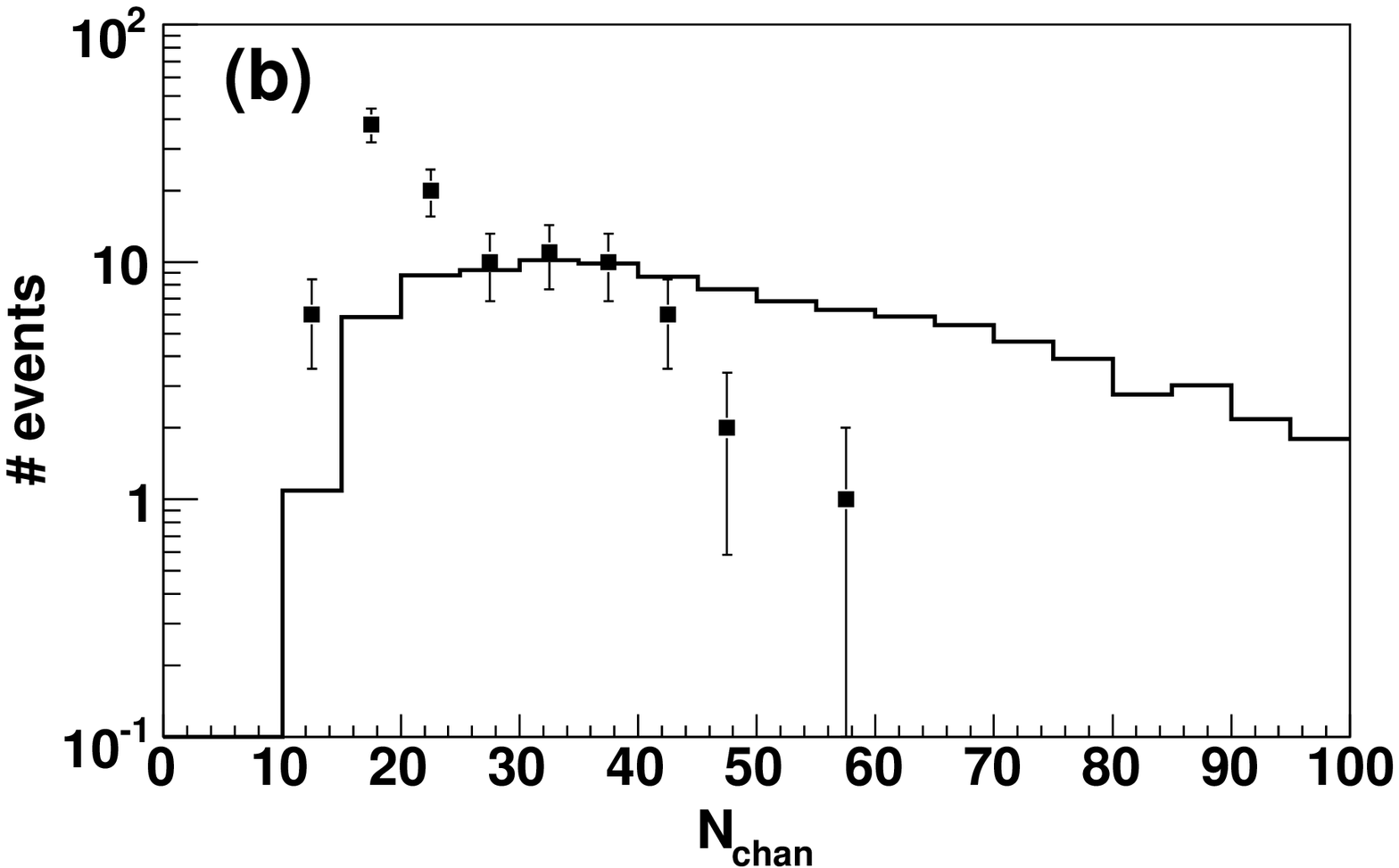}
\caption{(a) Likelihood ratio distributions for background-only
(off-time) data sets (solid line) and for the same background data sets
with one (dashed line) and two (dotted line) signal events injected.
(b) $N_\mathrm{chan}$ distributions for off-time data (squares), and
signal Monte Carlo (solid line) after final cuts. Data events are
selected within a radius of $20^\circ$ around the GRB position. The
Monte Carlo has been normalized to the number of events in
data.\label{fig:nChan}}
\end{figure}

To determine whether a given data set is compatible with the
background-only hypothesis, a large number of background data sets for
the on-time window are generated from the 2006 nine-string data by
randomizing the track times while taking into account the downtime of
the detector. For each of these data sets, the $\ln({\cal R})$ value is
calculated, yielding the distribution shown in Figure~\ref{fig:nChan}(a).
The probability for a data set to be compatible with background is
given by the fraction of background data sets with an equal or larger
$\ln({\cal R})$ value ($p$-value). The sensitivity of the IceCube
detector to neutrinos from GRB\,080319B is determined by injecting
different numbers of Monte Carlo signal events into the generated
randomized background data sets and calculating the resulting
$p$-value (see Figure~\ref{fig:nChan}(a)).

GRBs are expected to generate a substantially harder neutrino energy
spectrum than that of atmospheric neutrinos. A detector quantity
closely related to the neutrino energy is the number of DOMs
(channels) with detected photons, $N_\mathrm{chan}$. This quantity is
used to enhance the sensitivity to a possible
signal. Figure~\ref{fig:nChan}(b) shows the $N_\mathrm{chan}$
distribution of off-time data events within a cone with $20^\circ$
radius centered on the GRB position (no events remain in the on-time
data set) together with a signal Monte Carlo which has been normalized
to the number of data events. Combining this information with the
$p$-value from the $\ln({\cal R})$ distribution increases the
detection probability. The potential for discovering a muon neutrino
fluence as calculated in Section~\ref{sec:nuspec}
($\Gamma_\mathrm{jet} = 300$) with a significance of $4\sigma$ or
larger is about 6\%.

\subsection{Results and systematic uncertainties}\label{sec:results}
\begin{figure}[t]
\epsscale{1}
\plottwo{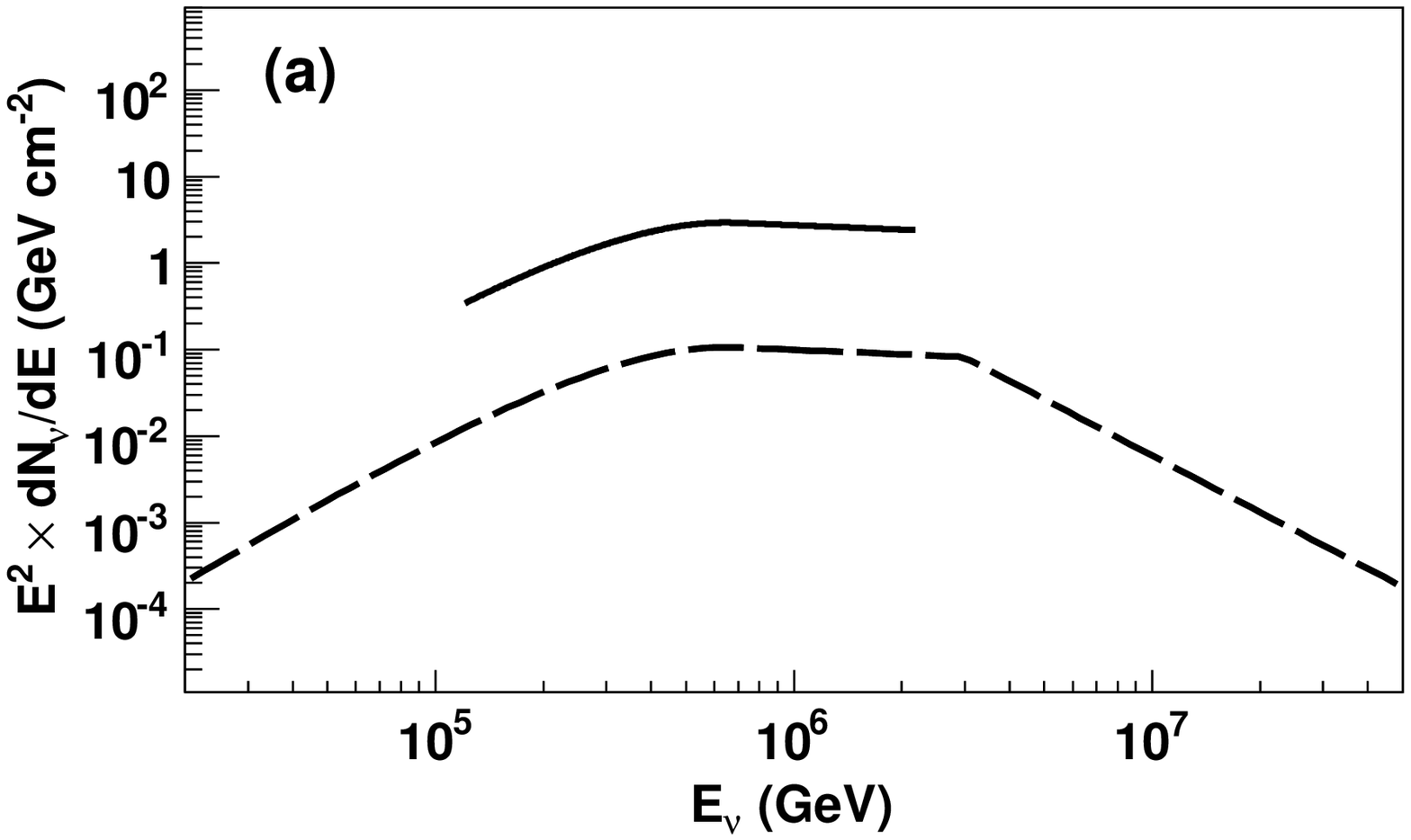}{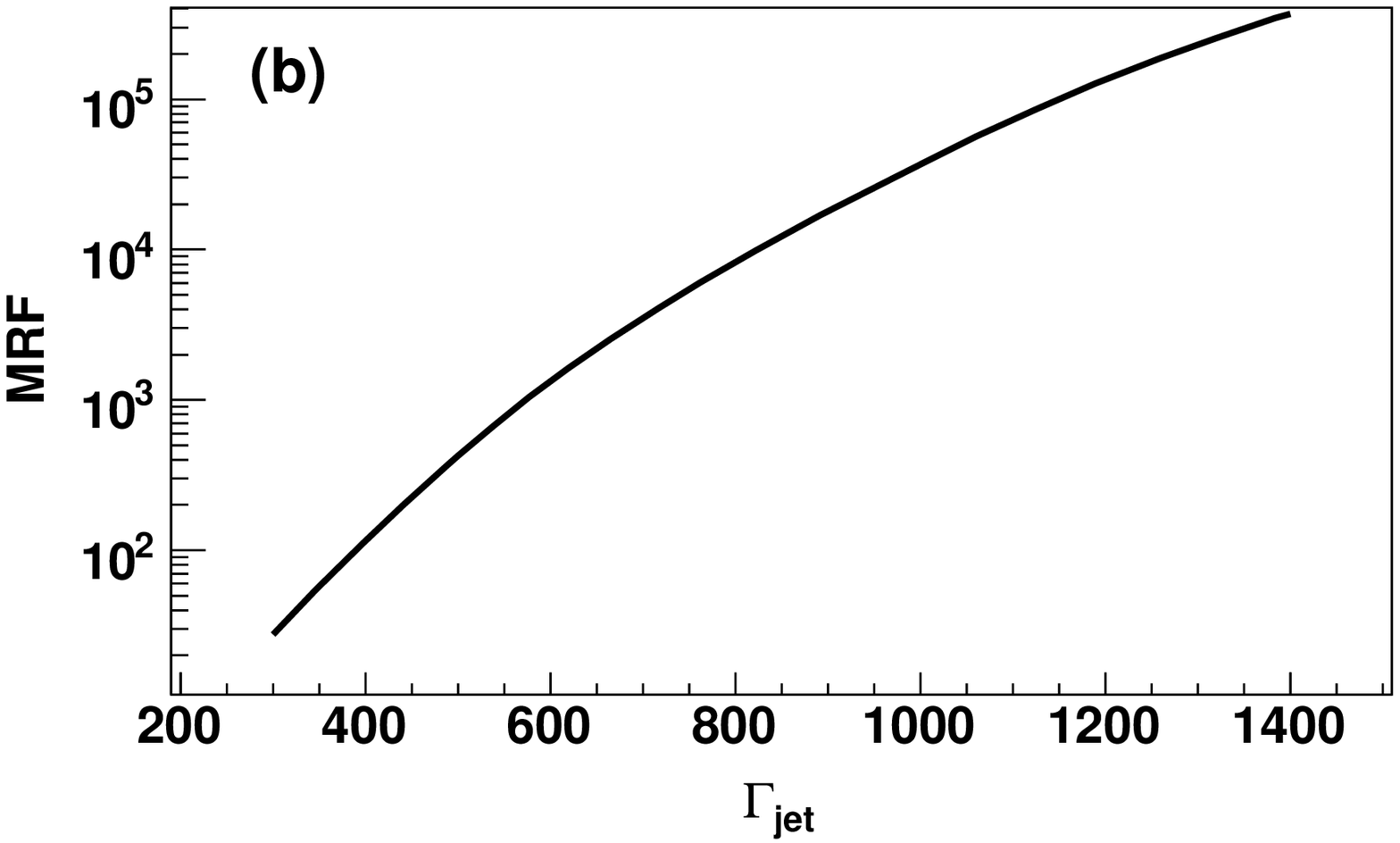}
\caption{(a) Ninety percent CL upper limit on the
fluence from GRB\,080319B (solid line) with respect to the calculated
neutrino fluence for a bulk Lorentz boost of the jet of
$\Gamma_\mathrm{jet} = 300$ (dashed line). (b) Ratio between upper
limit and calculated fluence spectrum (MRF) as a function of
$\Gamma_\mathrm{jet}$.\label{fig:limit}}
\end{figure}

After all parameters of the analysis have been fixed, the on-time data are
unblinded and analyzed with respect to two time windows\footnote{The
time windows define the flat part of the signal time PDFs.}. The
shorter one (from $T_0 - 3.8\,$s to $T_0 + 62.2\,$s) corresponds to
the immediate emission time of the $\gamma$-rays, whereas the second
(from $T_0 - 173\,$s to $T_0 + 130\,$s) covers the whole time range
with IceCube data (see Figure~\ref{fig:gammacurve}). No significant
deviation from the background-only hypothesis was found in either of
the two time windows. In both cases, the unbinned likelihood method
yields $\ln ({\cal R}(\langle \hat{n}_s \rangle)) = 0$ with $\langle
\hat{n}_s \rangle = 0$ as the best estimate for the number of signal
events. The resulting Neyman 90\% CL upper limit
\citep{ptrsl:a236:333,pl:b667:1} on the number of signal events in the
short time window is 2.7, i.e., injecting signal events according to a
Poisson distribution with mean 2.7 into the on-time data yields $\ln
({\cal R}(\langle \hat{n}_s
\rangle))>0$ in 90\% of cases. From the neutrino fluence calculated
in Section~\ref{sec:nuspec} for $\Gamma_\mathrm{jet} = 300$ we expect
to see 0.1 events. This results in a model rejection factor (MRF;
\citet{app:19:393}) of 27 which is defined as the ratio between the
upper limit on the number of signal events and the expected number of
signal events. The corresponding upper limit on the muon neutrino
fluence is plotted in Figure~\ref{fig:limit}(a) in the energy range
containing 90\% of the detected signal events. Integrating the fluence
over this energy range yields an upper limit of $9.5\times
10^{-3}\,\mathrm{erg}\,\mathrm{cm}^{-2}$. This upper limit is slightly
conservative as we do not consider the effect of $\nu_\tau$ from
GRBs. Tau neutrinos might manifest themselves as $\tau$ tracks (which
can travel a substantial distance at PeV energies) or as muons from
tau decays\footnote{Electron neutrinos do not contribute to the signal
as the resulting electron immediately produces an electromagnetic
cascade.}. The impact of a larger bulk Lorentz boost on the MRF is
displayed in Figure~\ref{fig:limit}(b). In addition, the MRFs for
$\Gamma_\mathrm{jet} = 500$ and 1400 are listed in Table~\ref{tab3}.

The effects of systematic errors on the result, described below in
detail, were investigated by varying simulation parameters and
repeating the full analysis. The quadratic sum of all systematic
errors is ($+17\%$,$-4\%$). The main sources of systematic
uncertainty are as follows.

\emph{Ice simulation:} Inaccuracies in the ice simulation can lead to a
wrong estimate of the efficiency of the detector to neutrinos from the
GRB. In order to estimate the size of this effect, the analysis was
repeated using a modified ice simulation. In this simulation, the DOM
efficiency was altered as a function of depth according to the
differences observed between data and Monte Carlo in the DOM
occupancy, effectively making the ice clearer for depths with
above-average transparencies and dirtier for depths with below-average
transparencies. This leads to an increase of the fluence upper limit
of 16\%, which is included as a one-sided systematic error.

\emph{DOM efficiency:} Uncertainties in the efficiency of the optical
modules in the photon detection lead to an uncertainty in the number
of expected events from the GRB. Varying the efficiency by $\pm10\%$
changes the upper limit by $\pm4\%$.

\emph{Background rate:} Even after optimized cuts, the data set is
dominated by misreconstructed downgoing atmospheric muons. The rate
varies throughout the year due to changes in the density profile of
the atmosphere at high altitude above the South Pole by about 10\%
around the mean value. As the number of events after cuts in the data
set during the GRB is too low, the rate at the time of the GRB was
determined using a 1\,hr data set recorded with the same DAQ settings
about a week later. In order to account for potential differences, the
background data rate was varied by $\pm 10\%$. This results in a shift
of the upper limit of less than $\pm1\%$.

\section{Comparison with other results}
In \citet{pr:d78:101302}, the author calculates neutrino spectra for a
burst like GRB\,080319B with different models. The expected fluences
in the best cases are below our optimistic scenario. Therefore, our
limits do not constraint these models.

In \citet{apj:697:730}, the Super-Kamiokande collaboration reports on
upper limits on the neutrino fluence from GRB\,080319B. They analyzed
their data in different energy ranges between $\sim\!100$\,MeV and
$\sim\!1$\,TeV. Their strongest fluence limits come from the
high-energy \emph{upmu} data set and amount to $16\,\mathrm{cm}^{-2}$
($22\,\mathrm{cm}^{-2}$) at 90\% CL for (anti) muon neutrinos
assuming an $E^{-2}$ spectrum. In comparison, IceCube reaches an upper
limit of $1.1\times10^{-5}\,\mathrm{cm}^{-2}$ in the energy range
$E_\nu = 120$\,TeV--2.2\,PeV. A direct comparison of the results is
difficult as the Super-Kamiokande paper does not contain information
about the neutrino energy integration range used in the calculation of
the limits. Assuming a lower integration boundary of $E_{\nu,1} =
10$\,GeV and an upper boundary $E_{\nu,2} \gg E_{\nu,1}$ yields a
sensitivity for an $E^{-2}$ spectrum that is about 200 times worse
than that of IceCube (for equal fluxes of muon and anti muon
neutrinos).

Along with high-energy neutrinos which origin from the decay of
charged pions high-energy $\gamma$-ray photons are produced in the
decay of simultaneously generated neutral pions. In addition,
high-energy photons are produced in inverse Compton scattering of
synchrotron photons by accelerated electrons
\citep{arXiv:0810.0520}. In contrast to neutrinos, the flux of
high-energy $\gamma$-ray photons at the Earth is significantly reduced due
to the large optical depths for photon--photon pair production inside
the source for not too large jet Lorentz factors $\Gamma_\mathrm{jet}
\lesssim 800$ \citep{arXiv:0810.0520}. In addition, high-energy photons above
100\,GeV are absorbed on the extragalactic background light if they
travel distances $z \gtrsim 0.5$. Observations with air-Cherenkov
telescopes like H.E.S.S.\ \citep{apj:690:1068,arXiv:0902.1561} or MAGIC
\citep{apj:667:358,proc:aip:magic:1} are also hampered by the fact
that usually it takes more than 50\,s (MAGIC) or 100\,s (H.E.S.S.)
from the observation of a GRB by a satellite to the start of data
taking with these telescopes. Therefore, the prompt emission window is
only partially covered or not at all. Milagro
\citep{apj:604:L25,apj:666:361} as an air shower array observed the
visible sky continuously. However, it was mostly sensitive to energies
above 1\,TeV. Up to now, there has been no definitive detection of
very high energy $\gamma$-ray emission from GRBs. Milagrito
\citep{apj:533:l119,apj:583:824} and the HEGRA AIROBICC array
\citep{aa:337:43} reported evidence at the $3\sigma$ level for
high-energy $\gamma$-ray emission from GRB\,970417A ($E_\gamma >
650$\,GeV) and GRB\,920925C ($E_\gamma > 20$\,TeV),
respectively. However, subsequent searches for high-energy $\gamma$-ray
emission from GRBs did not find similar signals. For GRB\,080319B, no
high-energy $\gamma$-ray data exist (MAGIC, which has a threshold
below 100\,GeV, was not able to observe the burst as it was already
dawning \citep{privcom:lorenz:2008:1}). Current flux predictions are near or
below the sensitivity of current instruments
\citep{arXiv:0810.0520}, where the predicted fluxes in the energy
range below $\sim\!100$\,TeV are dominated by the leptonic emission
component in most scenarios. Therefore, the upper limits from
high-energy $\gamma$-ray observations do not constrain the neutrino
flux in our model or the computed upper limit.

\section{Conclusions and outlook}
We used the IceCube neutrino telescope to search for high-energy muon
neutrinos from GRB\,080319B, one of the most spectacular and well
measured GRBs ever observed. Based on the fireball model of GRBs and
the measured $\gamma$-ray fluence we calculated the expected neutrino
spectrum for different jet bulk Lorentz boosts
$\Gamma_\mathrm{jet}$. After applying quality cuts to suppress
misreconstructed atmospheric muons a mean number of 0.1 signal events
is expected for the optimistic case of $\Gamma_\mathrm{jet} = 300$
(for other $\Gamma_\mathrm{jet}$ see Table~\ref{tab3}) in IceCube,
which was running in a nine-string configuration. The data were
analyzed with an unbinned log-likelihood method utilizing the
directional and temporal distance of reconstructed tracks to the
GRB. The sensitivity to a potential GRB signal was enhanced by
evaluating energy information. No deviation from the background-only
hypothesis was found either in a small time window covering the
immediate $\gamma$-ray emission time or an extended window of
300\,s. This results in a 90\% CL upper limit on the muon neutrino
fluence from GRB\,080319B within the short time window of
$9.5\times10^{-3}\,\mathrm{erg}\,\mathrm{cm}^{-2}$ in the energy range
between 120\,TeV and 2.2\,PeV which contains 90\% of the expected
signal events. The corresponding ratio between the upper limit and the
calculated GRB spectrum (MRF) is 27. Its stability with respect to
systematic uncertainties in the analysis is estimated to be
($+17\%$,$-4\%$). The upper limit does not allow us to impose
constraints on GRB parameters within the fireball model.

In its final configuration with 80 strings the expected number of
detected events in IceCube from a bright GRB like GRB\,080319B is
${\cal O}(1)$, rendering the individual analysis of these rare GRBs
highly interesting also in the future. Using the large number of GRBs
observed by the \emph{Swift} and \emph{Fermi} (formerly \emph{GLAST}) satellites, the
growing IceCube detector will also soon be able to probe the
Waxman-Bahcall or similar GRB fluxes and in the case of a
non-detection set stringent limits.

\acknowledgments
{\small We acknowledge the support from the following agencies:
U.S. National Science Foundation-Office of Polar Program,
U.S. National Science Foundation-Physics Division, University of
Wisconsin Alumni Research Foundation, U.S. Department of Energy, and
National Energy Research Scientific Computing Center, the Louisiana
Optical Network Initiative (LONI) grid computing resources; Swedish
Research Council, Swedish Polar Research Secretariat, and Knut and
Alice Wallenberg Foundation, Sweden; German Ministry for Education and
Research (BMBF), Deutsche Forschungsgemeinschaft (DFG), Germany; Fund
for Scientific Research (FNRS-FWO), Flanders Institute to encourage
scientific and technological research in industry (IWT), Belgian
Federal Science Policy Office (Belspo); the Netherlands Organisation
for Scientific Research (NWO); M.~Ribordy acknowledges the support of
the SNF (Switzerland); A.~Kappes and A.~Gro{\ss} acknowledge support
by the EU Marie Curie OIF Program; M. Stamatikos is supported by an
NPP Fellowship at NASA-GSFC administered by ORA.} 

\begin{deluxetable}{lccc}
\tablecaption{Neutrino Spectrum Parameters According to the Fireball
Model for GRB\,080319B for Different Bulk Lorentz Factors of the Jet
Together with the Expected Mean Number of Events in the Detector and
the Model Rejection Factor Obtained from the Analysis\label{tab3}}
\tabletypesize{\scriptsize}
\tablewidth{0pt}
\tablehead{
\colhead{Parameter $\backslash$ bulk Lorentz factor} & \colhead{$\Gamma_\mathrm{jet} = 300$} & \colhead{$\Gamma_\mathrm{jet} = 500$} & \colhead{$\Gamma_\mathrm{jet} = 1400$} 
}
\startdata
1st break energy, $\epsilon_1$ & 260\,TeV   & 710\,TeV  & 5.6\,PeV \\
2nd break energy, $\epsilon_2$ & 3\,PeV  & 23\,PeV & 1.4\,EeV \\
1st index, $\alpha_\nu$        & 0.5        & 0.5       & 0.5   \\
2nd index, $\beta_\nu$         & 2.17      & 2.17     & 2.17 \\
3rd index, $\gamma_\nu$        & 4.17      & 4.17     & 4.17 \\
Fluence at 1st break energy & $7.3\times10^{-13}\,\mathrm{GeV}^{-1}\,\mathrm{cm}^{-2}$ 
	                    & $1.4\times10^{-14}\,\mathrm{GeV}^{-1}\,\mathrm{cm}^{-2}$ 
                            & $2.9\times10^{-18}\,\mathrm{GeV}^{-1}\,\mathrm{cm}^{-2}$ \\
\tableline
Expected events in IceCube & 0.10  & $8.6\times10^{-3}$ & $9.6\times10^{-6}$ \\
MRF                  & 27        & 420      & $3.7\times10^5$ \\
\enddata
\end{deluxetable}

\clearpage

\appendix
\section{Equations used in the calculation of the neutrino spectrum}\label{app:nu_spec}
\begin{align}
F_\gamma(E_\gamma) &=
\frac{\mathrm{d}N(E_\gamma)}{\mathrm{d}E_\gamma}\\
	&= f_\gamma \times
             \begin{cases}
	      \left(\frac{E_\gamma}{\mathrm{MeV}}\right)^{-\alpha_\gamma} \exp\left(-\frac{E_\gamma}{\epsilon_\gamma}\right)
                & \text{for $E_\gamma < \epsilon_\gamma \left( \beta_\gamma - \alpha_\gamma \right)$ } \\
	      \left(\frac{E_\gamma}{\mathrm{MeV}}\right)^{-\beta_\gamma}
              \left[\left(\beta_\gamma-\alpha_\gamma\right)\frac{\epsilon_\gamma}{\mathrm{MeV}}\right]^{\beta_\gamma-\alpha_\gamma} 
              \exp\left(\alpha_\gamma-\beta_\gamma\right)
	        & \text{for $E_\gamma \ge\epsilon_\gamma \left( \beta_\gamma - \alpha_\gamma \right)$}
             \end{cases}
	     \label{eq:gamma_spec}\\[2mm]
{\cal F}_\gamma &=
		\int_{20\,\mathrm{keV}}^{7\,\mathrm{MeV}} \mathrm{d}E_\gamma
                \ E_\gamma F_\gamma(E_\gamma) \label{eq:gammaIntegral}\\[5mm]
F_\nu(E_\nu) &= \frac{\mathrm{d}N(E_\nu)}{\mathrm{d}E_\nu}\\
             &= f_\nu \times
             \begin{cases}
             \left(\frac{E_\nu}{\mathrm{GeV}}\right)^{-\alpha_\nu} \exp\left(-\frac{E_\nu}{\epsilon_1}\right)\hspace{2cm}
                \\ \hspace{6.6cm} \text{for $E_\nu < \epsilon_1 \left( \beta_\nu - \alpha_\nu \right)$} \\
             \left(\frac{E_\nu}{\mathrm{GeV}}\right)^{-\beta_\nu}
                \left[\left(\beta_\nu-\alpha_\nu\right)\frac{\epsilon_1}{\mathrm{GeV}}\right]^{\beta_\nu-\alpha_\nu}
	        \exp\left(\alpha_\nu-\beta_\nu\right)
                \\ \hspace{6.6cm} \text{for $\epsilon_1 \left( \beta_\nu - \alpha_\nu \right) \le E_\nu <
                \epsilon_2$} \\
             \left(\frac{E_\nu}{\epsilon_2}\right)^{-\gamma_\nu} 
                \left[\left(\beta_\nu-\alpha_\nu\right)\frac{\epsilon_1}{\mathrm{GeV}}\right]^{\beta_\nu-\alpha_\nu}
	        \exp\left(\alpha_\nu-\beta_\nu\right)
                \left(\frac{\epsilon_2}{\mathrm{GeV}}\right)^{-\beta_\nu}
                \\ \hspace{6.6cm} \text{for $E_\nu \ge \epsilon_2$}
            \end{cases}
	     \label{eq:nu_spec}\\[2mm]
\epsilon_1 &= 7\times 10^{5}\,\mathrm{GeV} \ \frac{1}{(1+z)^2}
              \left(\frac{\Gamma_\mathrm{jet}}{10^{2.5}}\right)^2
              \left( \frac{\mathrm{MeV}}{\epsilon_\gamma} \right)
\label{eq:epsilon1}\\[2mm]
\epsilon_2 &= 10^7 \, \mathrm{GeV} \ \frac{1}{1+z} \ 
              \sqrt{\frac{\epsilon_e}{\epsilon_B}} \ 
              \left(\frac{\Gamma_\mathrm{jet}}{10^{2.5}}\right)^4
              \left( \frac{t_\mathrm{var}}{0.01\,\mathrm{s}} \right) \ 
              \sqrt{\frac{10^{52}\,\mathrm{erg\,s}^{-1}}{L_\gamma^\mathrm{iso}}}
\label{eq:epsilon2}\\[2mm]
\alpha_\nu &= 3 - \beta_\gamma \quad , \quad \beta_\nu \ = \ 3 -
\alpha_\gamma \quad , \quad \gamma_\nu = \beta_\nu + 2 \\[2mm]
\frac{\Delta R}{\lambda_{p\gamma}} &=
	\left( \frac{L_\gamma^\mathrm{iso}}{10^{52}\,\mathrm{erg\,s}^{-1}} \right) \ 
	\left( \frac{0.01\,\mathrm{s}}{t_\mathrm{var}} \right) \ 
	\left(\frac{10^{2.5}}{\Gamma_\mathrm{jet}}\right)^4 \ 
        \left( \frac{\mathrm{MeV}}{\epsilon_\gamma} \right)
\label{eq:nint}\\[2mm]
\int_{0}^\infty \mathrm{d}E_\nu &\ E_\nu F_\nu(E_\nu) = 
	\frac{1}{8} \ 
        \frac{1}{f_e} \ 
	\left(1 - (1-\langle x_{p\rightarrow\pi} \rangle)^{\Delta R/\lambda_{p\gamma}}\right)
	\int_0^\infty \mathrm{d}E_\gamma \ E_\gamma F_\gamma(E_\gamma)
\label{eq:nuIntegral}
\end{align}
The parameters of the $\gamma$-ray spectrum $F_\gamma(E_\gamma)$ are
the break energy $\epsilon_\gamma$ and the spectrum indices before and
after the break $\alpha_\gamma$ and $\beta_\gamma$, respectively. The
quantity ${\cal F}_\gamma$ is the measured $\gamma$-ray fluence and
$z$ the redshift of the GRB. The parameters of the neutrino spectrum
$F_\nu(E_\nu)$ are the two break energies, $\epsilon_1$ and
$\epsilon_2$, and the corresponding spectral indices $\alpha_\nu$,
$\beta_\nu$, and $\gamma_\nu$. The expression $1 - (1-\langle
x_{p\rightarrow\pi}
\rangle)^{\Delta R/\lambda_{p\gamma}}$ in Equation~(\ref{eq:nuIntegral}) estimates
the overall fraction of the proton energy going into pions from the
size of the shock, $\Delta R$, and the mean free path of a proton for
photomeson interactions, $\lambda_{p\gamma}$. Here, $\langle
x_{p\rightarrow\pi} \rangle = 0.2$ is the average fraction of proton
energy transferred to a pion in a single interaction. The expression
ensures that the transferred energy fraction is $\le 1$. The variables
$f_\gamma$ and $f_\nu$ are obtained from the integrals of
Equations~(\ref{eq:gammaIntegral}) and (\ref{eq:nuIntegral}). The
isotropic equivalent luminosity, $L_\gamma^\mathrm{iso}$, is given by
the isotropic equivalent energy released in $\gamma$-rays,
$E_\gamma^\mathrm{iso}$, divided by the burst duration. The
calculations are insensitive to the beaming effect caused by a narrow
opening angle of the jet ($0^\circ\!\!.\,4$ for GRB\,080319B according to
\citet{nature:455:183}) as all formulae contain the isotropic luminosity
in conjunction with a $4\pi$ shell geometry, i.e., effectively use
luminosity per steradian. For example, the target photon density used
to calculate $\lambda_{p\gamma}$ is given by $n_\gamma \propto
L_\gamma^\mathrm{iso} / 4\pi R^2$.

\end{document}